\documentclass[USenglish,a4paper,11pt]{article}
\pdfoutput=1

\input{Header/style.sty}
\title{Quantum Advantage with Sign-Free Proofs: \\ An Oracle Separation between StoqMA and MA}

\author{Yassine Hamoudi\thanks{\texttt{ys.hamoudi@gmail.com}}\qquad\; Yvan Le Borgne\thanks{\texttt{borgne@labri.fr}}\qquad\; Shrinidhi Teganahally Sridhara\thanks{\texttt{shrinidhi.teganahally-sridhara@u-bordeaux.fr}} \\ [12pt] Université de Bordeaux, CNRS, LaBRI, France}

\date{\today}

\begin{document}

\maketitle

\begin{abstract}
  Several landmark oracle results, including~\cite{Wat00c,GNT25c,BHNZ26c}, establish strict separations along the chain $\mathsf{MA} \subseteq \mathsf{QCMA} \subseteq \mathsf{QMA}$, showing that quantum verification and quantum witnesses can each provide additional power over classical proof systems. These separations, however, use substantial coherent quantum computation, often through the quantum Fourier transform. We prove that a minimal quantum resource already suffices to outperform every classical verifier. Namely, we show the separation
  \[\mathsf{MA} \subsetneq \mathsf{StoqMA}\]
relative to a classical oracle. The class $\mathsf{StoqMA}$, introduced by Bravyi, Bessen, and Terhal~\cite{BBT06p}, restricts the verifier to classical reversible computation, followed by a single Hadamard-basis measurement. Its power is closely tied to the complexity of \emph{stoquastic} Hamiltonians, a sign-problem-free and seemingly more tractable class of quantum Hamiltonians.

Our separating problem is to determine whether a sparse high-girth graph, given via an adjacency-list oracle, has a large or small Perron--Frobenius eigenvalue. This amounts to a ground-energy estimation task for a related stoquastic Hamiltonian. Our construction builds on a line of work on graph exploration problems with exponential quantum speedups, from welded trees~\cite{CCD03c} to recent constructions~\cite{GHV21c,HLBS26p}.
\end{abstract}


\section{Introduction}
Understanding generalisations of the complexity class $\NP$ is key to identifying the right quantum analogue of classical proof systems (see~\cite{Gha24j}). One natural classical generalisation is $\MA$, where the verifier is allowed to use randomness and fail with bounded probability. Among the many quantum generalisations, one is $\StoqMA$, introduced by Bravyi, Bessen and Terhal~\cite{BBT06p}, in which the verifier is a very restricted quantum computer. It may use $\ket{0}$ and $\ket{+}$ ancillas, apply polynomially many classical reversible gates, and finish with a single Hadamard-basis measurement. Estimating the ground-state energy of stoquastic Hamiltonians is $\StoqMA$-complete when they are local~\cite{BBT06p}, or even just sparse~\cite{GR26p}. Here, stoquastic (or sign-problem-free) Hamiltonians are those whose off-diagonal entries are non-positive in the computational basis, allowing their ground states to be chosen with non-negative amplitudes. Despite its simplicity, this verifier model captures an important regime of Hamiltonian complexity. For example, Cubitt and Montanaro~\cite{CM16j} give an exhaustive classification of $2$-local Hamiltonian problems into four complexity classes: $\cP$, $\NP$-complete, $\StoqMA$-complete and $\QMA$-complete. While $\StoqMA$ is contained in $\QMA$, it has also been shown to sit in between $\MA$ and $\AM$, the two randomized extensions of $\NP$~\cite{BDOT08j,BBT06p}. The question of whether quantum proofs can be strictly more powerful than classical ones is central to quantum verification. $\StoqMA$ provides a natural setting for this question: does its minimal quantum power already suffice to outperform $\MA$?

At least three reasons suggest that $\MA$ might be equal to $\StoqMA$. First, its gate set is nearly classical. Second, classical methods such as quantum Monte Carlo algorithms appear to work efficiently on ground-state problems for stoquastic Hamiltonians~\cite{BT10j,Bra15j}. Third,~\cite{AGL25j} showed that if $\StoqMA$ admits error reduction, as most other proof-system classes do, then $\MA = \StoqMA$. In this work, we give evidence against this belief by showing that there exists a classical oracle relative to which $\MA \subsetneq \StoqMA$, so that a very restricted quantum verifier can solve a problem that no classical verifier with a classical witness can solve efficiently.

\begin{theorem}[\sc Oracle separation of $\StoqMA$ and $\MA$]\label{thm:main}
    There exists a classical oracle $\mcl{O}$ such that $\StoqMA^{\mcl{O}} \neq \MA^{\mcl{O}}$.
\end{theorem}

Our separating problem is to distinguish whether a sparse, high-girth graph on $2^m$ vertices has a large or small top eigenvalue $\lambda_{\mathrm{max}}(A)$ (its Perron--Frobenius eigenvalue) of its adjacency matrix~$A$. The graph is given through its adjacency-list oracle after a uniformly random relabeling of its vertices. Since $A$ is entrywise non-negative, $H=-A$ is a stoquastic Hamiltonian with ground energy $-\lambda_{\mathrm{max}}(A)$. The problem is therefore a ground-energy distinction problem for a stoquastic Hamiltonian, which is exactly what a $\StoqMA$ verifier is built to handle.

We construct such an oracle by building on and extending prior work by Gilyén, Hastings, and Vazirani~\cite{GHV21c}, as well as our own prior work~\cite{HLBS26p}. The two graph families have polynomially separated top eigenvalues. The $\yes$ instance is a single graph $G$ and for every small seed set $U$, we construct a $\no$-instance $F(U)$ that is indistinguishable from $G$ when exploration starts from $U$. A $\StoqMA$ verifier receives the top eigenvector itself as a quantum witness and verifies its energy using a simple Hadamard test, a more elementary procedure than quantum phase estimation. An $\MA$ verifier, in contrast, is essentially limited to local graph exploration and cannot detect the global spectral difference using a classical witness and $\poly(m)$ queries.

\medskip 

Oracle separations are a central tool for studying the power of quantum verification and quantum proofs. The choice of oracle model is crucial, as illustrated by the separations between $\QCMA$ and $\QMA$ (where a quantum verifier receives a classical or quantum proof). Aaronson and Kuperberg~\cite{AK07j} first established this separation relative to \emph{unitary oracles}. Such oracles are particularly powerful because they provide black-box access to an arbitrary unitary transformation~$U$. Fefferman and Kimmel~\cite{FK18c} later obtained the separation using randomized \emph{in-place permutation oracles}, whose queries permute computational-basis states as $\ket{x} \mapsto \ket{\pi(x)}$. After a long line of refinements, two recent works~\cite{BHNZ26c,BHV26p} finally established the separation in the standard \emph{classical-oracle} model~\cite{Deu85j}, in which a function $f$ is accessed through the reversible XOR map $\ket{x,y}\mapsto \ket{x,y\oplus f(x)}$. The choice of oracle model is particularly important for $\StoqMA$, whose verifier is restricted to classical reversible gates. Allowing arbitrary unitary oracles would violate this restriction. Even phase oracles (obtained from classical oracles by phase kickback) introduce signs. We therefore work with classical XOR oracles, which preserve the sign-free structure of the verifier.


\subsection{Relation to prior work}

\paragraph{Expander-mixing based separations.}

Earlier partial progress on the $\QMA$ vs.\ $\QCMA$ question came from subspace and subset verification~\cite{AK07j,FK18c}, Yamakawa--Zhandry-based oracles~\cite{LLPY24c,BK24c}, and expander-mixing problems over graph oracles~\cite{Lut11p,NN24j,BFM23c,LMY25c}. These results either require special oracle access or rely on substantial coherent quantum computation, often through the quantum Fourier transform. The expander-mixing candidates are closest in spirit to our graph problem, but a fundamental difference poses an obstacle to verification in $\StoqMA$. They use regular graphs, hence the top eigenvalue is trivial to compute. The difficulty lies instead in testing the second-largest eigenvalue of the adjacency matrix $A$ (equivalently the second-lowest eigenvalue of $-A$). Our problem instead concerns the top eigenvalue itself (whose negative is the ground-state energy of~$-A$), precisely the quantity a $\StoqMA$ verifier is suited to certify. This reflects a broader distinction for stoquastic Hamiltonians: ground-energy estimation is $\StoqMA$-complete, whereas highest-energy estimation is $\QMA$-complete~\cite{JGL10j}.

\paragraph{Other oracle separations.}

Several other oracle separations have appeared recently, including $\QMA_1$ vs.\ $\QCMA$~\cite{MPR26c} and $\QMA(2)$ vs.\ $\QMA$~\cite{BGH+26p,GMT26p}. Here, the subscript~$1$ denotes perfect completeness, while $\QMA(2)$ allows two unentangled quantum proofs. The result of~\cite{MPR26c} also separates $\StoqMA_1$ from $\QCMA$, as we sketch in \Cref{App:preimage-stoqma}, but uses an in-place oracle. The construction in~\cite{GMT26p} is also close to our setting: its hard instances are signed graphs accessed through a classical adjacency-list oracle, as in the earlier constructions of~\cite{NN24j,LMY25c}. However, its $\yes$ and $\no$ instances are both frustration-free. Thus, unlike in our problem, the two cases are distinguished not by the lowest energy over all states, but by the lowest energy attainable by product states.

\paragraph{Adiabatic and guided stoquastic problems.}

Our work is most directly related to two earlier oracle separations for stoquastic problems: that of Gilyén, Hastings, and Vazirani~\cite{GHV21c} in the adiabatic setting, and our earlier work~\cite{HLBS26p} in the dequantized guided ground-state setting. The present setting differs in how auxiliary information is supplied. Here it is an untrusted proof, whereas~\cite{HLBS26p} assumes a trusted guide as part of the input, and \cite{GHV21c} uses no witness. The graph constructions in all three works inherit ideas from welded trees~\cite{CCD03c} but require substantial changes for their respective settings. Although the decision problem studied in~\cite{HLBS26p} belongs to $\StoqMA$, we show in \Cref{App:earlier-problem-np} that it also belongs to $\NP$, requiring a different construction for an oracle separation. Our $\yes$ instances use the same graph construction as~\cite{HLBS26p}, combining a high-degree, high-girth base graph with the self-similar decorations of~\cite{GHV21c}. Here, however, the base graph need not be an expander. Our new $\no$ instances build on a notion of closure that has appeared in earlier work (e.g.,~\cite{CMM09c,BCK15c}) and that we study further here.


\subsection{Proof overview}

We outline the proof of \Cref{thm:main}, which gives an oracle separation between $\StoqMA$ and $\MA$.

\paragraph{Distinguishing problem and graph families.}
Our separation is based on estimating the top eigenvalue of a sparse graph given oracle access to its adjacency list. We call this the Graph Perron-Eigenvalue Estimation (GPE) problem. We construct two families of graphs on exponentially many vertices ($2^{\Theta(n^3\log^2{n})}$) whose top eigenvalues are polynomially separated. To make the problem hard for classical algorithms, the vertex names are hidden by a uniformly random relabeling, so an algorithm can learn about the graph only by exploring it \emph{locally} from vertices it already knows.

\begin{figure}[t]
  \begin{center}
    \input{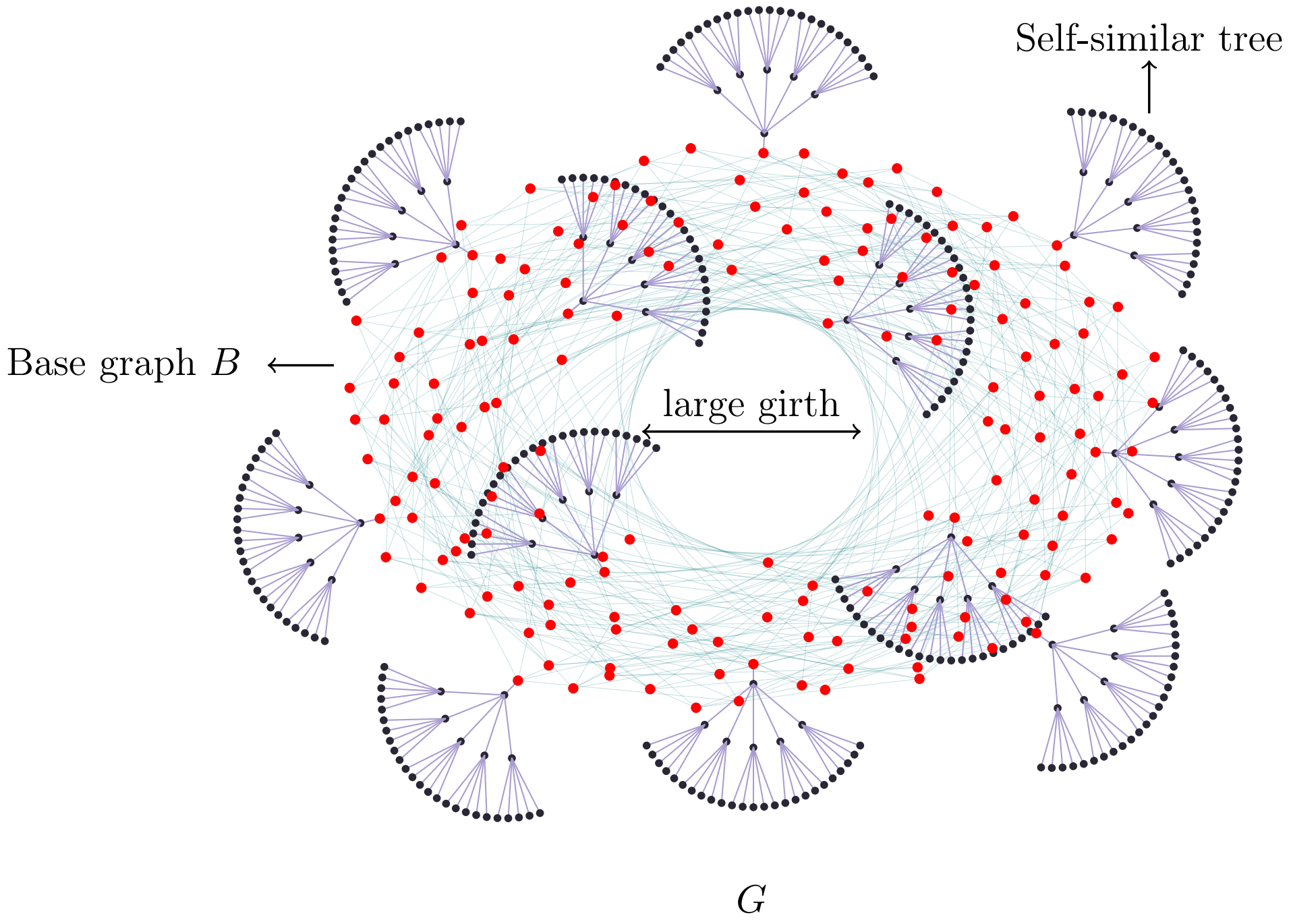}
    \caption{Pictorial representation of the $\yes$ graph $G$. The high-girth, regular base graph $\base$ is shown with red vertices and green edges, with the self-similar trees attached to its vertices shown in purple; for simplicity, we show them on only some of the vertices, with only one copy per vertex, and depict each self-similar tree as a simple perfect tree. The actual construction is more complicated: each tree is obtained by attaching shorter-depth trees to a perfect tree (see~\Cref{Def:maingraph,Def:selfsim}).}
  \label{Fig:maingraph}
  \end{center}
\end{figure}

\emph{The $\yes$ family} consists, for each $n$, of all labelings of a single graph $G$ (\Cref{Def:maingraph} and \Cref{Fig:maingraph}). To construct $G$, we start from an $n$-regular \emph{base graph} of girth $g=\poly(n)$ (\Cref{lem:base}), similar to that of~\cite{HLBS26p}, and attach copies of \emph{self-similar trees} (\Cref{Def:selfsim}) to every vertex of the base graph, similar to those studied in~\cite{Has21j,GHV21c}. Because of its girth, the base graph looks like a tree locally, and the shape of the attached trees mimics this local structure (\Cref{prop:iso}), so that locally the trees are hard to tell apart from the base graph. The resulting graph $G$ has maximum degree less than $2n$, and since it contains the $n$-regular base graph, whose top eigenvalue is $n$, we have~$\lambda_{\max}(G)\geq n$.

\begin{figure}[t]
  \begin{center}
    \input{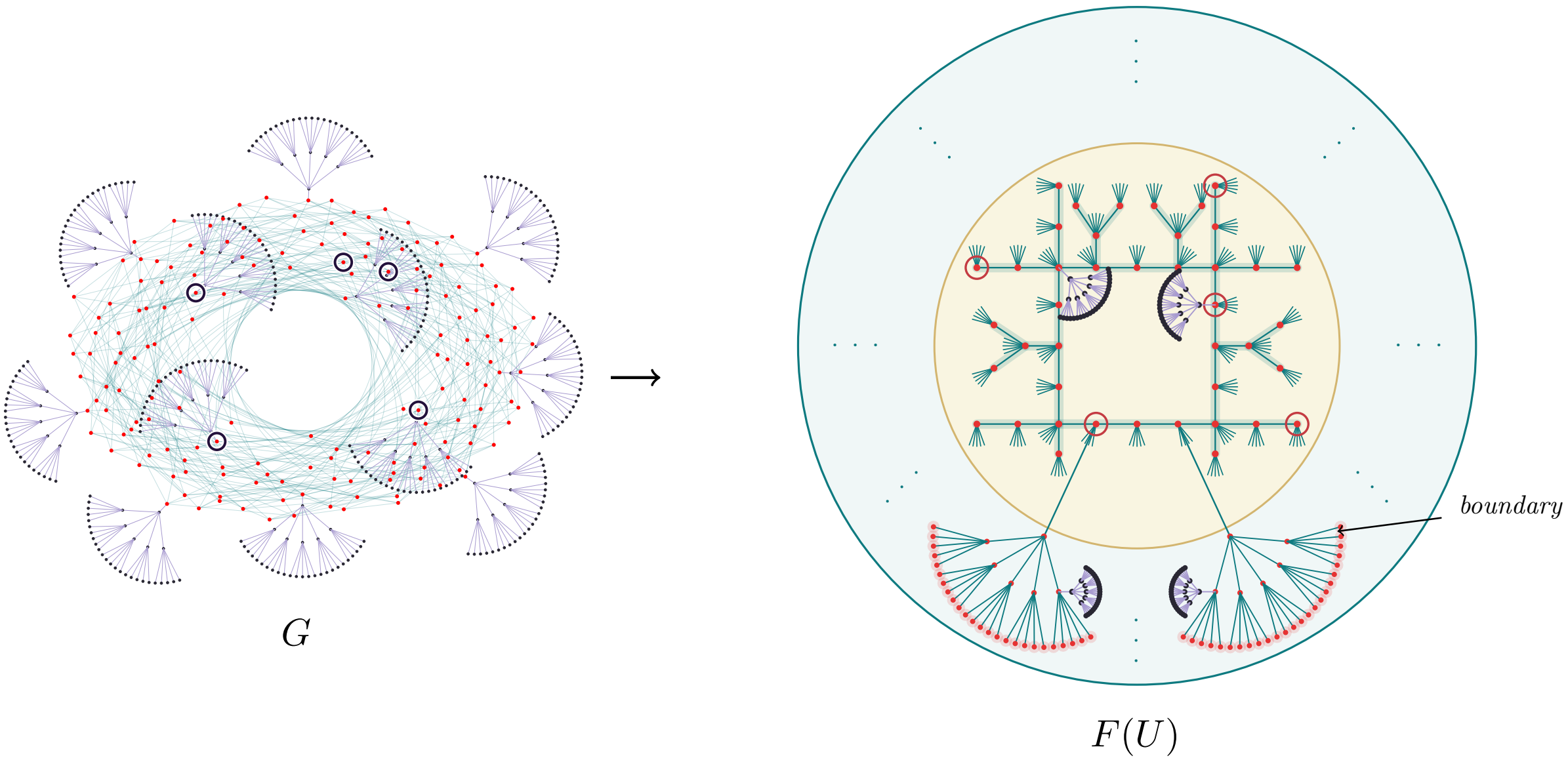}
    \caption{Pictorial representation of the $\no$ graph $F(U)$, obtained as a subgraph of $G$ using $U$. The vertices of $U$ are circled in both graphs. In $F(U)$, on the right, the yellow region marks the closure of $U$, and the short paths added between vertices to form the closure are highlighted. For two vertices, the subgraph reachable via an edge leaving the closure is shown together with its attached purple trees. These subgraphs are pairwise disjoint, and each is isomorphic to the same fixed self-similar tree; the \emph{boundary vertices} (highlighted in pink) correspond exactly to a particular set of its leaves. As in the previous figure, self-similar trees are shown attached to only a few vertices, and only one copy is drawn (see~\Cref{Def:closuregraph} and \Cref{prop:iso} for a detailed description).}
  \label{Fig:nograph}
  \end{center}
\end{figure}

\emph{The $\no$ family} is designed to have a small top eigenvalue while looking like $G$ to an algorithm exploring locally, even when given a small set of vertices as starting points (we later derive such a set from the witness via a bit-fixing technique). Concretely, for each \emph{seed set} $U$ of at most~$2^{\sqrt{n}}$ vertices, a graph~$F(U)$ is obtained by truncating~$G$ around $U$~(\Cref{Def:closuregraph} and \Cref{Fig:nograph}). The truncation must not be recognisable by local exploration starting from $U$, so we keep everything within a large distance of $U$ and truncate only far away from it. Since the same trees hang off every base vertex, only the base graph distinguishes one place of $G$ from another, so we truncate by distance in the base graph (a tree vertex counts as the base vertex it hangs off).

We build $F(U)$ as follows. We start from $U$ and add all vertices on short paths between its vertices, repeating until nothing new is added; the result, a set of base vertices, is the \emph{closure} of $U$ (\Cref{def:closure}). We then keep everything within a large (polynomial) distance of the closure and delete the rest. By the high girth, for every edge going out of the closure, all vertices reachable from the closure via this edge induce a subgraph that is isomorphic to a particular self-similar tree~(\Cref{prop:iso}), and these subgraphs are pairwise disjoint. The truncation is chosen so that the outermost retained base vertices, called the \emph{boundary}, correspond exactly to a particular set of leaves in these trees. These are the only places where $F(U)$ stops while~$G$ continues.

These graphs inherit the high girth $g$ of $G$. Moreover, each $F(U)$ is \emph{$2$-degenerate} (\Cref{lem:2-degenerate}): it contains no subgraph of minimum degree at least $3$ (equivalently, it can be reduced to the empty graph by repeatedly removing a vertex of degree at most $2$). Indeed, by the Moore bound (\Cref{lem:avg-moore}), any subgraph of the closure with minimum degree at least $3$ would have at least~$2^{\Omega(g)}$ vertices. However, the closure is only polynomially larger than its seed set, containing at most $\poly(n)\cdot|U|$ vertices (\Cref{lem:closure-size}). For our choice of $g$ and $\abs{U}\leq 2^{\sqrt{n}}$, this is too small to contain such a subgraph. The rest of~$F(U)$ consists of trees attached to the closure, which preserves $2$-degeneracy. In contrast, $G$ has degeneracy exactly~$n$ since the base graph is $n$-regular. A central consequence of being sparse and $2$-degenerate is that the spectral norm is small, namely $\lambda_{\max}(F(U))\leq 4\sqrt{n}$~(\Cref{lem:small-spec-deg}). This gives the spectral separation from the $\yes$ instance.

\paragraph{$\StoqMA$ protocol.}
The GPE problem is easily shown to lie in $\StoqMA$ because the negative of the adjacency matrix $-A$ is a sparse stoquastic Hamiltonian~\cite{GR26p}. In \Cref{prop:gpe-stoqma}, we give a direct proof using the Hadamard test. For any graph of maximum degree at most~$d$, our verifier uses $\poly(n)$ adjacency-list queries and accepts any witness $\ket{\psi}$ with probability $1/2+\langle\psi|A|\psi\rangle/(2d)$. It therefore accepts a top eigenvector of $G$ ($\yes$ case) with probability at least $3/4$, while accepting every witness for $F(U)$ ($\no$ case) with probability at most $1/2+1/\sqrt{n}$.

\paragraph{Classical hardness when the witness is a set of vertices.}
As a first step toward showing that GPE is not in $\MA$, we consider the simplest kind of witness: one that just reveals the names of a set $U$ of at most $2^{\sqrt n}$ vertices, so that the algorithm knows where $U$ is, while all other vertices stay hidden behind random names. We show that this does not help: no classical algorithm making polynomially many queries can distinguish $G$ from the $\no$ graph $F(U)$ built around $U$ with non-negligible advantage, even though it knows what $U$ is.

An algorithm can tell $G$ from $F(U)$ only by finding a place where they differ. Since all vertices outside $U$ are hidden behind random names, a classical algorithm cannot jump to a vertex it has not reached: except with negligible probability, it only explores the part of the graph connected to $U$ by following edges (this is standard~\cite{CCD03c}; see~\Cref{lem:conncomp}). Throughout such an exploration, the two graphs look the same until the algorithm reaches a \emph{boundary} vertex, which is a leaf of $F(U)$ but has further neighbors in $G$~(\Cref{prop:iso}). Hence the algorithm must find a path from some vertex in $U$ to the boundary.

Why is this hard? Recall that by construction, the part of $F(U)$ reachable after taking an edge leaving the closure of $U$ is a copy of the same self-similar tree, and the boundary vertices correspond to a particular set of leaves of this tree~(\Cref{prop:iso} and \Cref{Fig:nograph}). So, starting from a vertex of the closure, the algorithm has to walk down the self-similar tree attached to it, all the way to one of these particular leaves, without being able to tell which way leads there. The self-similar trees are designed to hide these leaves from any classical algorithm~\cite{GHV21c,HLBS26p}: a polynomial-query algorithm finds such a path with probability at most $2^{-2n\log n}$ (\Cref{Lem:hardss}).

So far we considered a path starting from one fixed vertex of the closure, that is, a descent down a single self-similar tree, and showed that finding it is hard. To reach the boundary, the algorithm needs such a path from \emph{some} vertex of the closure. Since the trees hanging off distinct closure vertices are disjoint and independently labelled, exploring one reveals nothing about the others, so \Cref{Lem:hardss} bounds the success probability in each of them by $2^{-2n\log n}$ regardless of the algorithm's history; a union bound over all of them then applies. The closure is relatively small: it has at most $\poly(n)\cdot|U|$ vertices (\Cref{lem:closure-size}). Altogether, the probability of reaching the boundary is thus at most $q\cdot\poly(n)\cdot|U|\cdot 2^{-2n\log n}$, where $q$ is the number of queries, which stays negligible even for $|U|$ as large as $2^{\sqrt n}$.

\paragraph{From vertices to arbitrary witnesses.}
An $\MA$ witness need not be a set of vertices: it is an arbitrary bit-string of polynomial length, and it may depend on the entire graph, including how its vertices are named. We would like to show that such a string cannot do more than reveal the location of a small set of vertices. If this were true, the proof would conclude as follows. Take a $\yes$ instance $G$ and a witness that makes the verifier accept it. By the claim, this witness is as good as revealing the location of some small set $U$. But an algorithm that only knows where $U$ is cannot tell $G$ from $F(U)$ by the previous paragraph, so the verifier would also accept the $\no$ instance $F(U)$ with the same witness, contradicting soundness.

To make this precise, we use a bit-fixing technique for random permutations~\cite{CDG18c} (see~\Cref{lem:lab-fixing}). Informally, it says that polynomially many bits of information about a uniformly random relabeling can be replaced, up to a small error, by revealing the names of a few vertices while keeping all other names uniformly random. (The lemma speaks about queries to the relabeling rather than to the graph, which only costs a polynomial factor in the number of queries.) We apply it with up to $2^{\sqrt n}$ revealed vertices: this is enough to make the error $\poly(n)/2^{\sqrt n}$ negligible, and it is still within the size of seed sets handled before. The witness then keeps the verifier's acceptance probability on $G$ at least $2/3-2^{-\sqrt n/2}$, and once the witness is fixed the argument above goes through (\Cref{lem:arb-witness}).

Finally, a standard diagonalisation argument~\cite{Zha25c,NN24j}, adapted in~\Cref{thm:main}, lifts this to a relativized separation. We build a single oracle $\mcl{O}$ whose $n$-th block is the adjacency-list oracle of a GPE instance on $2^{\Theta(n^3\log^2 n)}$ vertices from either the $\yes$ or the $\no$ family, chosen so that the $i$-th candidate $\MA$ verifier errs on its block. Since the $\StoqMA$ protocol above solves GPE for every such oracle, this gives a relativized separation of $\StoqMA$ and $\MA$.

\section{Preliminaries}
\subsection{Complexity classes}
\label{sec:comp-class}

A promise problem $\mcl{L} = (\lyes,\lno)$ consists of disjoint subsets $\lyes, \lno \subseteq \{0,1\}^*$. We now give the definitions of the two complexity classes considered in this paper.

\begin{definition}[\sc $\MA$ and $\MA^{\mcl{O}}$]
    \label{def:ma}
    A promise problem $\mcl{L} = (\lyes,\lno)$ is in the complexity class $\MA$ if and only if there exists a polynomial $p(\cdot)$ and a probabilistic polynomial-time verifier~$\mcl{V}(x,w)$ where $\mcl{V}$ takes as input a bit string $x \in \{0,1\}^{*}$ and a witness $w \in \{0,1\}^{p(|x|)}$ and outputs either~$0$ (reject) or $1$ (accept) such that:
    \begin{itemize}
        \item[] \textbf{Completeness:} If $x\in \lyes$, then there exists $w$ such that $\pr*{\mcl{V}(x,w)=1} \geq 2/3$.
        \item[]  \ \ \ \ \ \textbf{Soundness:} If $x\in \lno$, then for any $w$, $\pr*{\mcl{V}(x,w)=1} \leq 1/3$.
    \end{itemize}

    The oracle version $\MA^{\mcl{O}}$ is defined in the same way, for an oracle $\mcl{O}:\{0,1\}^*\rightarrow\{0,1\}$, except that $\lyes^{\mcl{O}}$ and $\lno^{\mcl{O}}$ may depend on $\mcl{O}$ and the verifier $\mcl{V}^{\mcl{O}}(x,w)$ has oracle access to $\mcl{O}$, which it may query at most polynomially many times.
\end{definition}

The completeness and soundness parameters in the definition of $\MA$ can be chosen arbitrarily, provided that they are inverse-polynomially separated from each other and from $0$ and $1$. This is unlike $\StoqMA$, for which gap amplification remains an open problem. We recall the definition introduced in~\cite{BBT06p}.

\begin{definition}[\sc $\StoqMA$, \cite{BBT06p}] \label{def:stoqma}
    A \emph{stoquastic verifier} is a tuple $\mcl{V} = (m, m_w, m_0, m_+, U)$, where $m$ is the number of input bits, $m_w$ is the number of input witness qubits, $m_0$ is the number of input ancillas in the state $\ket{0}$, $m_+$ is the number of input ancillas in the state~$\ket{+}$, and $U$ is a quantum circuit on $m+m_w + m_0 + m_+$ qubits with $X$, $\mathrm{CNOT}$, and Toffoli gates. The acceptance probability of a stoquastic verifier $\mcl{V}$  on input string $x\in\{0,1\}^m$ and witness state $\ket{\psi}\in (\mathbb{C}^2)^{\otimes m_w}$ is defined as $\Pr[\mcl{V}(x, \ket{\psi})=1] = \braket{\psi_{\rm inp}}{U^\dagger \Pi_{\rm out} U}{\psi_{\rm inp}}$, where $\ket{\psi_{\rm inp}} = \ket{x}\otimes\ket{\psi}\otimes\ket{0}^{\otimes m_0} \otimes \ket{+}^{\otimes m_+}$ is the initial state and $\Pi_{\rm out} = {\op{+}{+}}_1\otimes \id_\text{else}$ projects the first qubit onto the state $\ket{+}$.

    \smallskip
    A promise problem $\mcl L = (\lyes, \lno)$ belongs to $\StoqMA(a,b)$ if there exists a polynomial-time uniform family of stoquastic verifiers, such that for any input $x\in\mcl L_{\yes} \cup \mcl L_{\no}$ of $m$ bits, the corresponding stoquastic verifier~$\mcl{V}$ uses at most $p_1(m)$ qubits and $p_2(m)$ gates where both $p_1$ and $p_2$ are polynomials and obeys the following:

    \begin{itemize}
        \item[] \textbf{Completeness}: If $x \in \lyes$, then there is some $\ket{\psi}$ such that $\pr{\mcl{V}(x, \ket{\psi})=1} \geq a(m)$.
        \item[] \ \ \ \ \ \textbf{Soundness}: If $x \in \lno$, then for any $\ket{\psi}$, $\Pr[\mcl{V}(x, \ket{\psi})=1] \leq b(m)$.
    \end{itemize}

    We define $\StoqMA$ as the union of $\StoqMA(a,b)$ over all polynomial-time computable threshold functions $a,b$ satisfying $1/2 \leq b(m) <a(m) \leq 1$ and $a(m) - b(m)\geq 1/\poly(m)$.

    To define the oracle version $\StoqMA^{\mcl{O}}(a,b)$, fix an oracle $\mcl{O}: \{0,1\}^{*} \rightarrow \{0,1\}$. The sets $\lyes^{\mcl{O}}$ and $\lno^{\mcl{O}}$ may depend on $\mcl{O}$ and the verifier $\mcl{V}^{\mcl{O}}(x,\ket{\psi})$ has oracle access to $\mcl{O}$, which it may query at most polynomially many times via query gates $\ket{y}\ket{c}\mapsto\ket{y}\ket{c \oplus \mcl{O}(y)}$.
\end{definition}

Both definitions extend to an oracle $\mcl{O}:\{0,1\}^*\to\{0,1\}^m$ with longer outputs, such as the adjacency-list oracle we define later in \Cref{Def:ora-graph}: replace each query by $m$ queries $\ket{y,j}\ket{c}\mapsto\ket{y,j}\ket{c\oplus\mcl{O}(y)_j}$ for $j\in[m]$, which increases the query count by only a polynomial factor. All our results are stated and proved with respect to this multi-bit oracle model.

\paragraph{XOR oracle.} We stress that the query gate must be the XOR oracle above and not the phase-flip oracle $\ket{y}\ket{c}\mapsto(-1)^{c \cdot \mcl{O}(y)}\ket{y}\ket{c}$. For general quantum verifiers the two are equivalent via Hadamard gates on the target qubit, but these are not available to a $\StoqMA$ verifier. 


\subsection{Graph-theoretic notions}
\label{sec:graphs}

We consider simple undirected graphs $G=(V,E)$, where $V$ and $E$ denote the vertex and edge sets, respectively. A graph $G'=(V',E')$ is a \emph{subgraph} of $G$ if $V'\subseteq V$ and $E'\subseteq E$, with every edge in $E'$ having both endpoints in $V'$. The subgraph of $G$ \emph{induced} by $V'\subseteq V$ is denoted by $G[V']=(V',E_{\mid V'})$, where $E_{\mid V'}=\{\{u,v\}\in E:u,v\in V'\}$. The degree of a vertex $v\in V$, denoted by $\deg(v)$, is the number of vertices adjacent to $v$, namely $\deg(v)=|\{u\in V:\{u,v\}\in E\}|$.

\begin{definition}[\sc $c$-degenerate graph, \cite{LW70j}]
    \label{def:c-degenerate}
    A graph $G$ of $n$ vertices is called $c$-degenerate if every non-empty subgraph of $G$ contains at least one vertex of degree at most $c$.
\end{definition}

\begin{lemma}[\sc Spectral norm of $c$-degenerate graphs, \cite{Hay06c}]
    \label{lem:degen-spec}
    The spectral norm of the adjacency matrix of a $c$-degenerate graph $G$ with maximum degree $d$ is at most $2\sqrt{cd}$.
\end{lemma}

\begin{proof}
    By $c$-degeneracy, we can repeatedly remove a vertex whose degree in the remaining graph is at most $c$. Orient each edge from the endpoint removed earlier to the endpoint removed later. The resulting directed adjacency matrix $B$ has at most $c$ nonzero entries per row and at most $d$ per column, and satisfies $A_G=B+B^\top$. Thus $\norm{B}_\infty\leq c$ and $\norm{B}_1\leq d$, giving $\norm{A_G}_2\leq 2\norm{B}_2\leq 2\sqrt{\norm{B}_1\norm{B}_\infty}\leq 2\sqrt{cd}$.
\end{proof}

A \emph{path} of length $\ell \ge 0$ in $G$ is a sequence of distinct vertices $u_0,u_1,\ldots,u_{\ell}$ such that $\{u_{i-1},u_i\}\in E$ for every $i\in\{1,\ldots,\ell\}$.
A \emph{cycle} of length $\ell \ge 3$ in $G$ is a sequence of distinct vertices $u_1,\ldots,u_{\ell}$ such that $\{u_i,u_{i+1}\}\in E$ for every $i\in\{1,\ldots,\ell-1\}$ and $\{u_{\ell},u_1\}\in E$.

\begin{definition}[\sc Girth of a graph]\label{def:girth}
    The girth of a graph $G$ denoted by $\girth(G)$ is the length of the shortest cycle in $G$.
\end{definition}

The following result gives a lower bound on the number of vertices of a graph in terms of its girth and average degree.

\begin{lemma}[\sc Average Moore bound, \cite{AHL02j}]
    \label{lem:avg-moore}
    The number of vertices $n$ in a graph of girth $g$ and average degree at least $d \geq 2$ satisfies: $n \geq (d-1)^{\lfloor \frac{(g-1)}{2} \rfloor}$.
\end{lemma}

In constructing the $\no$-graph family, we use a \emph{closure} operation that enlarges a vertex set until every short path between its vertices lies entirely within the set, and we prove a bound on the size of the result. Related closure operations have been studied in other contexts, such as LP integrality gaps and sum-of-squares lower bounds~\cite{CMM09c,BCK15c}.

\begin{definition}[\sc Closure of a set of vertices]
\label{def:closure}
    Let $G=(V,E)$ be a finite graph, let $r \geq 2$ be an integer, and let $U\subseteq V$. Starting from $U$, repeat the following step as long as possible: 
    \begin{itemize}
        \item[] If there is a path of length $\ell \in \{2,\dots,r\}$ whose two endpoints are distinct vertices of the current set $U$ and whose internal vertices all lie outside $U$, add its internal vertices to $U$.
    \end{itemize} 
    Since each step enlarges $U$, and $V$ is finite, the process terminates. The resulting set does not depend on the order in which the steps are performed; we call it the \emph{closure} of $U$ with parameter~$r$ and denote it by $\closure_{G,r}(U)$. We drop the subscript $G$ when the graph is clear from context.
\end{definition}

When the girth is large enough, the size of the closure scales linearly in the size of the initial set $\abs{U} = t$ and the distance parameter $r$. A similar bound was established in~\cite[Lemma 4.4]{BCK15c} under different hypotheses.

\begin{lemma}[\sc Size of closure]\label{lem:closure-size}
  Let $U \subseteq V$ be a set of vertices such that $\abs{U} = t \geq 1$ and~$r \geq 2$ be an integer. Suppose $G$ has girth at least $g \geq 4r\log(2rt) + 4$. Then the procedure of \Cref{def:closure} makes fewer than $2t$ additions and the final number of vertices is $\abs{\closure_r(U)} < 2rt$.
\end{lemma}

\begin{proof}
    We let $U_i$ denote the set $U$ after $i$ steps of adding paths as in \Cref{def:closure}, so $U_0 = U$ is the original set. Let $v_i = \abs{U_i}$ and $e_i$ be the number of edges of $G[U_i]$. We have $e_{i+1} - e_{i} \geq v_{i+1} - v_{i} + 1$, since the added path adds exactly $\ell-1$ vertices and at least $\ell$ edges in the induced subgraph (for some $\ell$). Applying this inequality $i$ times and using $v_0 = t$, $e_0 \geq 0$ gives $e_i - v_i \geq e_0 - v_0 + i \geq i - t$. Moreover, $v_i \leq t + i(r-1)$ since each step adds at most $r-1$ vertices.

    Suppose for contradiction that the process makes at least $2t$ steps. Then $v_{2t} < 2rt$ and $e_{2t} - v_{2t} \geq t$, so $G[U_{2t}]$ has average degree $d = 2e_{2t}/v_{2t} \geq 2 + 2t/v_{2t} > 2 + 1/r$. Since $G[U_{2t}]$ is a subgraph of $G$, it has girth at least $g$. By applying the Moore bound (\Cref{lem:avg-moore}), we get $v_{2t} \geq (d-1)^{\lfloor (g-1)/2 \rfloor} \geq (1 + 1/r)^{\lfloor (g-1)/2 \rfloor}$. Using $1 + x \geq \exp(\tfrac{x}{1+x})$ with $x = 1/r$, together with $\lfloor (g-1)/2 \rfloor \geq 2r\log(2rt)$ and $r + 1 \leq 2r$, we get $v_{2t} > 2rt$, contradicting $v_{2t} < 2rt$.

    Hence at most $2t-1$ paths are added, and the final set satisfies $|\closure_r(U)| < 2rt$.
\end{proof}

\section{Problem definition and setup}
In this section we introduce the problem that we show to lie in $\StoqMA$ but not in $\MA$. The task is to estimate the top eigenvalue of a graph given adjacency-list oracle access under a random labeling of its vertices. We first show that a stoquastic verifier solves this problem for every sparse graph, placing it in $\StoqMA$ (\Cref{prop:gpe-stoqma}). We then construct two families of sparse graphs whose top eigenvalues are separated by a polynomial gap in $m$, providing the $\yes$ and $\no$ instances (\Cref{sec:yes-graph,sec:no-graph}) used in our separation.

We access graphs through the standard adjacency-list oracle model, defined below.

\begin{definition}[\sc Oracle of a labelled graph]
  \label{Def:ora-graph}
  Let $G$ be a simple graph on $2^m$ vertices with maximum degree at most $d$, and let $\pi : V(G) \ra [2^m]$ be a permutation of its vertices, which we call a \emph{labeling}. 
  
  For $x \in [2^m]$ and $i \in [d]$, let $N(x)_i \in [2^m]$ be the label of the $i$th neighbor of $\pi^{-1}(x)$ (ordered by increasing label), or $x$ if this vertex has fewer than $i$ neighbors.
  \begin{itemize}
    \item \emph{(Classical oracle)} Given $(x,i) \in [2^m]\times[d]$, the oracle returns $\ora_{G,\pi}(x,i) = N(x)_i$.
    \item \emph{(Quantum oracle)} The oracle implements the unitary $\ora_{G,\pi}:\ket{x}\ket{i}\ket{z}\mapsto\ket{x}\ket{i}\ket{z\oplus N(x)_i}$.
  \end{itemize}
  For a labelled graph $(G,\pi)$, let $A_{G,\pi}:=P_\pi A_G P_\pi^{-1}$ be its permuted adjacency matrix, where~$P_\pi$ is the permutation matrix with $(P_\pi)_{x,u}=1$ if $\pi(u)=x$ and $0$ otherwise. It has the same spectrum as $A_G$, and if $\ket{\psi_G}$ is the top eigenvector of $A_G$ then $\psp =P_\pi\ket{\psi_G}$ is the top eigenvector of~$A_{G,\pi}$.

  When clear from the context, we may drop $G$ from the subscript.
\end{definition}


\subsection{The Graph Perron-Eigenvalue Estimation problem}
\label{sec:gpe}

Our separating problem asks whether a sparse graph, accessed through its adjacency-list oracle, has a large or small Perron eigenvalue.

\begin{definition}[\sc Graph Perron-Eigenvalue Estimation]
    \label{Def:gpe}
    The \emph{Graph Perron-Eigenvalue Estimation} (GPE) problem is defined as follows. We are given oracle access to the adjacency list of a labelled graph $G$ on $2^m$ vertices (see \Cref{Def:ora-graph}). The graph is assumed to be $\poly(m)$-sparse, meaning that $G$ has maximum degree at most $d = \poly(m)$. The top eigenvalue $\lambda_{\max}(G)$ of its adjacency matrix is promised to satisfy exactly one of the following:
    \[
      \textnormal{YES: } \lambda_{\max}(G) \geq a \qquad\textnormal{NO: } \lambda_{\max}(G) \leq b,
    \]
    where $a-b\geq 1/\poly(m)$. The goal is to output $1$ in the YES case and $0$ in the NO case.
\end{definition}

The matrix $-A_{G,\pi}$ is a stoquastic sparse Hamiltonian whose ground-state energy is the negative of the top eigenvalue of $A_{G,\pi}$. Estimating this energy can be done in $\StoqMA$ given the adjacency-list oracle, by a recent result of Grilo and Rozos~\cite{GR26p}. For completeness, we give a direct proof by constructing a stoquastic verifier that estimates the top-eigenvalue of the graph. 

\begin{lemma}[\sc Stoquastic verifier for GPE]
  \label{lem:stoq-verifier}
  Let $G$ be a simple graph on $2^m$ vertices of maximum degree at most $d = \poly(m)$. Then there
  exists a stoquastic verifier (\Cref{def:stoqma}) with access to the quantum
  oracle $\ora_{G,\pi}$ (\Cref{Def:ora-graph}) which, given any $m$-qubit state
  $\ket{\psi}$ as input, makes at most $\poly(m)$ queries to $\ora_{G,\pi}$ and
  accepts with probability
  \[
    \frac{1}{2} + \frac{\braket{\psi}{A_{G,\pi}}{\psi}}{2d}.
  \]
\end{lemma}

\begin{proof}
  For adjacent vertices with labels $x$ and $y$, let $i_{x\ra y} \in [d]$ denote the unique index satisfying $N(x)_{i_{x\ra y}} = y$. Since $G$ is undirected, the reverse position $i_{y\ra x}$ also exists, although the two positions need not coincide. Define, for any $x \in [2^m]$, $i \in [d]$, $b \in \rn$,
  \[
    U_{G,\pi}\ket{x,i,b}=
    \begin{cases}
      \ket{y,i_{y\ra x},b}, & \text{if $N(x)_i = y$ and $x \neq y$},\\
      \ket{x,i,1-b}, & \text{if $N(x)_i = x$}.
    \end{cases}
  \]
  This map is an involution on basis states and hence defines a unitary. Its action can be computed using $O(d)$ queries and $\poly(m)$ classical reversible gates and ancilla qubits initialized to $\ket{0}$.

  Let $\ket{u_d}=\frac{1}{\sqrt d}\sum_{i=1}^{d}\ket{i}=\ket{+}^{\otimes\log d}$ (we assume that $d$ is a power of two). Then $U_{G,\pi}$ satisfies the block-encoding identity,
  \begin{equation}\label{eq:compression}
    (\id\otimes\bra{u_d}\otimes\bra{0})\,U_{G,\pi}\,(\id\otimes\ket{u_d}\otimes\ket{0})=\frac{A_{G,\pi}}{d}.
  \end{equation}
  Indeed, each edge contributes $1/d$, while unoccupied positions flip the last qubit and therefore contribute zero. Given a witness $\ket{\psi}$, the verifier applies controlled-$U_{G,\pi}$ on $\ket{+}\ket{\psi}\ket{u_d}\ket{0}$, obtaining $\frac{1}{\sqrt{2}}\pt{\ket{0}\ket{\psi}\ket{u_d}\ket{0}+\ket{1}U_{G,\pi}(\ket{\psi}\ket{u_d}\ket{0})}$. It then measures the control qubit in the Hadamard basis and accepts on outcome $+$. By \Cref{eq:compression}, its acceptance probability is $\frac12+\frac12(\bra{\psi}\bra{u_d}\bra{0})U_{G,\pi}(\ket{\psi}\ket{u_d}\ket{0})=\frac12+\frac{1}{2d} \bra{\psi}A_{G,\pi}\ket{\psi}$.
\end{proof}

The verifier above converts the eigenvalue promise in GPE into a completeness--soundness gap, yielding the following containment in $\StoqMA$.

\begin{proposition}
  \label{prop:gpe-stoqma}
  The GPE problem (\Cref{Def:gpe}) belongs to $\StoqMA$.
\end{proposition}

\begin{proof}
  By \Cref{lem:stoq-verifier}, there is a stoquastic verifier that uses $\poly(m)$ queries to $\ora_{G,\pi}$ and accepts an $m$-qubit witness $\ket{\psi}$ with probability $\frac12+\frac{1}{2d}\bra{\psi}A_{G,\pi}\ket{\psi}$. If $G$ is a $\yes$ instance, then $\lambda_{\max}(G)\geq a$, so the top eigenstate $\psp$ of $A_{G,\pi}$ is accepted with probability at least $\frac12+\frac{a}{2d}$. If $G$ is a $\no$ instance, then $\lambda_{\max}(G)\leq b$, so every witness is accepted with probability at most $\frac12+\frac{b}{2d}$. Hence GPE belongs to $\StoqMA\big(\frac12+\frac{a}{2d},\frac12+\frac{b}{2d}\big)$. Since $d\leq\poly(m)$ and $a-b\geq1/\poly(m)$, the completeness--soundness gap is inverse polynomial, proving the claim.
\end{proof}


\subsection{The \texorpdfstring{$\yes$}{YES}-instances: High-degree, high-girth graphs}
\label{sec:yes-graph}

We describe the $\yes$ instance $G_n$ that underlies the separation. For each value of the parameter~$n$ there is a single such graph, so the $\yes$ instances form the sequence $(G_n)_n$. Throughout, $n$ is a free integer parameter, which
will later be chosen as $n=\poly(m)$. Following~\cite{HLBS26p}, the graph~$G_n$ is obtained from a high-degree, high-girth base graph (which, unlike in~\cite{HLBS26p}, need not be an expander) by decorating it with self-similar trees and adding dummy isolated vertices.

We first define the \emph{base graph} $\base_n$ using Dahan's construction~\cite{Dah14j}. This generalizes the Lubotzky--Phillips--Sarnak construction of Ramanujan graphs~\cite{LPS88j} to arbitrary sufficiently large degrees while preserving large girth.

\begin{lemma}[\sc High-degree, high-girth base graph $\base_n$~\cite{Dah14j}]
  \label{lem:base}
  For every sufficiently large integer $n$, there is an explicitly constructible connected simple $n$-regular graph $\base_n$ on $N_{\base}$ vertices satisfying
  \[
    2^{n^3\log^2 n} \leq N_{\base}\leq 2^{2n^3\log^2 n}
    \qquad\text{and}\qquad
    \girth(\base_n) \geq g_n:=100n^{5/2}\log n.
  \]
\end{lemma}

We then \emph{decorate} every vertex of $\base_n$ with copies of self-similar trees. This fractal-like decoration has appeared in constructions of stoquastic Hamiltonians~\cite{Has21j,GHV21c,HLBS26p}, and we use the version of~\cite{HLBS26p}.

\begin{definition}[\sc Self-similar trees $\treh_{n,k}$]
  \label{Def:selfsim}
  Let $n \geq 1$ and $1 \leq k \leq \sqrt{n}$. Define the degree and depth sequences $d_{n,k} = 2n - k\sqrt{n}$ and $\ell_{n,k} = k\cdot 10\, n^{3/2}\log n$.
  The \emph{perfect tree} $C_{n,k}$ is the rooted tree of depth $\ell_{n,k}$ in which every internal vertex has exactly $d_{n,k}-1$ children. Its leaves are called the \emph{$0$-level decorated leaves}.

  The self-similar tree $\treh_{n,k}$ is obtained from $C_{n,k}$ as follows. For each $j = 1,\dots,k-1$, attach to every internal vertex of $C_{n,k}$
  $d_{n,j}-d_{n,j+1}=\sqrt{n}$ new edges, and join the other endpoint of each new edge to the root of a fresh copy of $\treh_{n,j}$. The base case is
  $\treh_{n,1}=C_{n,1}$.
\end{definition}

Finally, we add sufficiently many isolated vertices that they form an overwhelming majority of the graph. This is a standard trick~\cite{CCD03c} to force classical algorithms to explore the nontrivial part of the graph by following edges from known vertices (\Cref{lem:conncomp}). The resulting graph~$G_n$, which serves as the $\yes$ instance, is defined below.

\begin{definition}[\sc Decorated base graph $G_n$]
  \label{Def:maingraph}
  Let $n \geq 1$ and $K = \sqrt{n}$. The graph $G_n$ is built in three steps.
  \begin{enumerate}
    \item Start from the $n$-regular base graph $\base_n$ on $2^{n^3\log^2 n} \leq N_{\base}\leq 2^{2n^3\log^2 n}$
    vertices with girth at least $g_n = 100n^{5/2}\log{n}$ (see \Cref{lem:base}).
    \item For each $k = 1,\dots, K-1$ and each vertex of $\base_n$, attach
    $d_{n,k} - d_{n,k+1} = \sqrt{n}$ new edges, joining the other endpoint of each
    to the root of a fresh copy of the self-similar tree $\treh_{n,k}$
    (see \Cref{Def:selfsim}).
    \item Add isolated vertices until the total number of vertices is
    $2^{15n^{3}\log^2{n}}$.
  \end{enumerate}
  The vertex set of $G_n$ is $V = V_{\base} \cup V_T \cup V_I$, where $V_{\base}$ consists of the \emph{base graph vertices} (the vertices of $\base_n$), $V_T$ of the \emph{tree vertices} (those in the copies of $\treh_{n,k}$), and $V_I$ of the \emph{isolated vertices}. When several graphs are in play, we write $V(G_n)=V$, and similarly for the other sets.
\end{definition}

To analyse $G_n$, we will need to relate the tree vertices to their attachment points in the base graph. We do so using a projection onto the base graph, which also induces a notion of distance between vertices. These notions will also be central to the construction and analysis of our $\no$ instances.

\begin{definition}[\sc Base projection and base distance]
  \label{Def:loc}
  For each $v\in V_{\base}\cup V_T$, let $\ep{v}\in V_{\base}$ denote its \emph{base projection}, defined as the unique base vertex closest to $v$ in $G_n$.
  
  The \emph{base distance} between $u,v \in V_{\base} \cup V_T$ is $\dist_{\base}(u,v)=\dist(\ep{u},\ep{v})$, where $\dist$ denotes graph distance in $\base$. Equivalently, this is the number of base graph edges on a shortest path from $u$ to $v$ in $G_n$. We set $\dist_{\base}(u,v)=+\infty$ if either vertex lies in $V_I$. For $U \subseteq V$, we define $\dist_{\base}(U,v)=\min_{u\in U}\dist_{\base}(u,v)$.
\end{definition}

Apart from the isolated vertices, the degree of every vertex in $G_n$ is either $d_{n,1}$ or $1$ (the degree-one vertices are exactly the leaves of the self-similar trees). The graph $G_n$ also has the same girth as $\base$. The top eigenvalue is at least $n$, so we can set $a = n$ in the associated GPE problem (\Cref{Def:gpe}).

\begin{lemma}[\sc Large Perron eigenvalue]
  \label{Fac:gn-spec}
  The top eigenvalue of~$G_n$ is at least $n$.
\end{lemma}

\begin{proof}
  The base graph $\base$ is an $n$-regular induced subgraph of $G_n$, so its top eigenvalue is exactly~$n$. The top eigenvalue of a graph is at least that of any of its subgraphs, so $\lambda_{\max}(G_n) \geq n$.
\end{proof}


\subsection{The \texorpdfstring{$\no$}{NO}-instances: Decorated closure graphs}
\label{sec:no-graph}

We construct the $\no$ family as a collection of subgraphs of $G_n$, one for each set $U$ of at most $2^{\sqrt{n}}$ vertices. The construction relies on the closure operation (\Cref{def:closure}) and is defined below.

\begin{definition}[\sc Decorated closure graph $\envl(U)$]
  \label{Def:closuregraph}
  Let $U \subseteq V$ be a subset of vertices of~$G_n$. The graph $\envl(U)$ is built in three steps.
  \begin{enumerate}
    \item Let $\br{U} = \closure_{2\ell_{n,K}+3}(\ep{U}) \subseteq V_{\base}$ be the closure vertices of the base projection of $U$ (\Cref{Def:loc,def:closure}).
    \item Let $W=\set{v\in V:\dist_{\base}(\br{U},v)\leq\ell_{n,K}}\cup\set{v\in V_{\base}:\dist_{\base}(\br{U},v)=\ell_{n,K}+1}$. The first set contains all \emph{base and tree} vertices of $G_n$ at base distance at most $\ell_{n,K}$ from $\br{U}$, while the second adds only \emph{base} vertices at base distance exactly $\ell_{n,K}+1$. We call the vertices in the second set the \emph{boundary vertices} of $U$, denoted by $\partial U$. 
    \item Define the graph $\envl(U) = (V, E_{\mid W})$ by extending the vertex set of $G_n[W]$ to $V$, with all added vertices isolated.
  \end{enumerate}
  The resulting graph $\envl(U)$ has the same vertex set as $G_n$ and exactly the edges of the induced subgraph $G_n[W]$. It is the member of the $\no$ family defined by $U$.
\end{definition}

The key combinatorial property of the closure graph $\envl(U)$ is that it is $2$-degenerate whenever the seed set~$U$ is sufficiently small.

\begin{lemma}[\sc $2$-degeneracy of closure]
  \label{lem:2-degenerate}
  If $\abs{U} \leq 2^{\sqrt{n}}$ then the graph $\envl(U)$ is $2$-degenerate.
\end{lemma}

\begin{proof}
  First, we show that $G_n[\br{U}]$ is $2$-degenerate. Suppose otherwise. By \Cref{def:c-degenerate}, it contains a non-empty subgraph $G'$ with minimum degree at least $3$. Since $G'$ has girth at least~$g_n$, the Moore bound (\Cref{lem:avg-moore}) gives $|V(G')| \geq (3-1)^{\lfloor(g_n-1)/2\rfloor}$. On the other hand, \Cref{lem:closure-size} gives $|V(G')| \leq \abs{\br{U}} < 2(2\ell_{n,K}+3)2^{\sqrt n} < 2^{\lfloor(g_n-1)/2\rfloor}$ for sufficiently large $n$, since $\ell_{n,K}=10n^2\log n$ and $g_n=100n^{5/2}\log n$. This contradiction proves that $G_n[\br{U}]$ is $2$-degenerate.

  Next, we show that the subgraph $G_n[W]$ remains $2$-degenerate. We claim that $G_n[W]$ is obtained from $G_n[\br{U}]$ by attaching trees at some vertices. Notice that such an operation preserves $2$-degeneracy. To prove the claim, take a breadth-first forest of the base vertices in $W$, rooted at~$\br{U}$. Each tree has depth at most $R:=\ell_{n,K}+1$. Any base edge belonging to neither this forest nor $G_n[\br{U}]$ would join vertices in the same tree, creating a cycle of length at most $2R+1$, or vertices in different trees, creating a path of length between $2$ and $2R+1$ between distinct vertices of $\br{U}$ with all internal vertices outside $\br{U}$. The former contradicts the girth bound $g_n>2R+1$, while the latter contradicts the closure property, since $2R+1=2\ell_{n,K}+3$. Hence every base edge outside $G_n[\br{U}]$ belongs to the breadth-first forest. Neither these edges nor the decoration edges create new cycles, proving the claim.

  Finally, $\envl(U)$ is $2$-degenerate as well since it adds no edges over $G_n[W]$.
\end{proof}

It follows from the $2$-degenerate property that the graph $\envl(U)$ has a small top eigenvalue.

\begin{proposition}[\sc Small Perron eigenvalue]
  \label{lem:small-spec-deg}
  If $\abs{U} \leq 2^{\sqrt{n}}$ then the top eigenvalue of~$\envl(U)$ is at most $4 \sqrt{n}$.
\end{proposition}

\begin{proof}
  This follows directly by applying \Cref{lem:degen-spec}, \Cref{lem:2-degenerate} and the fact that the maximum degree of the decorated closure graph is at most that in $G_n$, i.e., $d_{n,1} = 2n-\sqrt{n}$.
\end{proof}

It follows that the $\yes$ and $\no$ instances constructed above can be distinguished in $\StoqMA$.

\begin{corollary}[\sc $\StoqMA$ protocol for $G_n$ vs. $\envl(U)$]
  \label{lem:stoqma-family}
  There is a $\StoqMA$ protocol distinguishing $\yes$: $G_{n,\pi}$ from $\no$: $F_{n,\pi}(U)$, where $\pi$ is any labeling and $U \subset V$ any subset of at most $2^{\sqrt{n}}$ vertices.
\end{corollary}

\begin{proof}
  By \Cref{Fac:gn-spec} and \Cref{lem:small-spec-deg}, the top eigenvalues of the two graphs differ by at least $n - 4\sqrt{n}$. Moreover, both graphs have degree at most $2n-\sqrt{n}$. Hence, they represent valid instances of the Graph Perron-Eigenvalue Estimation problem (\Cref{Def:gpe}) and can be solved by a $\StoqMA$ verifier (\Cref{prop:gpe-stoqma}).
\end{proof}

\section{Lower bound when advice is a set of vertices}
In this section we show that no randomised algorithm can efficiently distinguish the two graph families even when given some subset of vertices as advice. More precisely, given any set of vertices $U$ and any fixed labeling of them as input, we prove that there is no randomised algorithm that can distinguish between $\main$ and $\envl(U)$ by making at most polynomially many queries to the graph oracle.

We first show that there is a partial isomorphism between $\main$ and $\envl(U)$ for any set of vertices $U$. Recall that both have the same vertex set $V$ and only differ in the edges adjacent to some vertices which is made precise in the following proposition.

\begin{proposition}[\sc Isomorphism up to the boundary]
  \label{prop:iso}
  Let $U\subseteq V$, and let $W$ and the boundary $\partial U\subseteq W$ be as in \Cref{Def:closuregraph}. Every vertex of $W \setminus \partial U$ has the same neighborhood in $\main$ and $\envl(U)$. The neighborhoods differ only at vertices of $\partial U \cup Y$, where $Y$ consists of the vertices outside $W$ that are non-isolated in $\main$. More precisely,
  \begin{enumerate}[label=(\alph*)]
    \item $N_{\envl(U)}(v)=N_{\main}(v)$ for every $v\in W\setminus\partial U$;
    \item $N_{\envl(U)}(v)=N_{\main}(v)\cap W\subsetneq N_{\main}(v)$ for every $v\in\partial U$;
    \item every vertex outside $W$ is isolated in $\envl(U)$.
  \end{enumerate}
  Consequently, for every labeling $\pi$ and every query $(x,i)$ with $\pi^{-1}(x)\in W\setminus\partial U$, $\ora_{\envl(U),\pi}(x,i)=\ora_{\main,\pi}(x,i)$.

  Moreover, the graph $\main[W]$ is isomorphic to the graph obtained from $\main[\br{U}]$ by attaching to each $u\in\br{U}$ as many copies of $\treh_{n,k}$ as it has in~$\main$ (for $k=1,\dots,K-1$), together with as many copies of~$\treh_{n,K}$ as the number of base graph neighbors of $u$ outside $\br{U}$. The isomorphism can be chosen so that it maps the boundary vertices, i.e., the base graph vertices at distance exactly $\ell_{n,K}+1$ from $\br{U}$ in~$\main$, bijectively onto the $0$-level leaves of the attached copies of~$\treh_{n,K}$.
\end{proposition}

\begin{figure}[t]
  \begin{center}
    \input{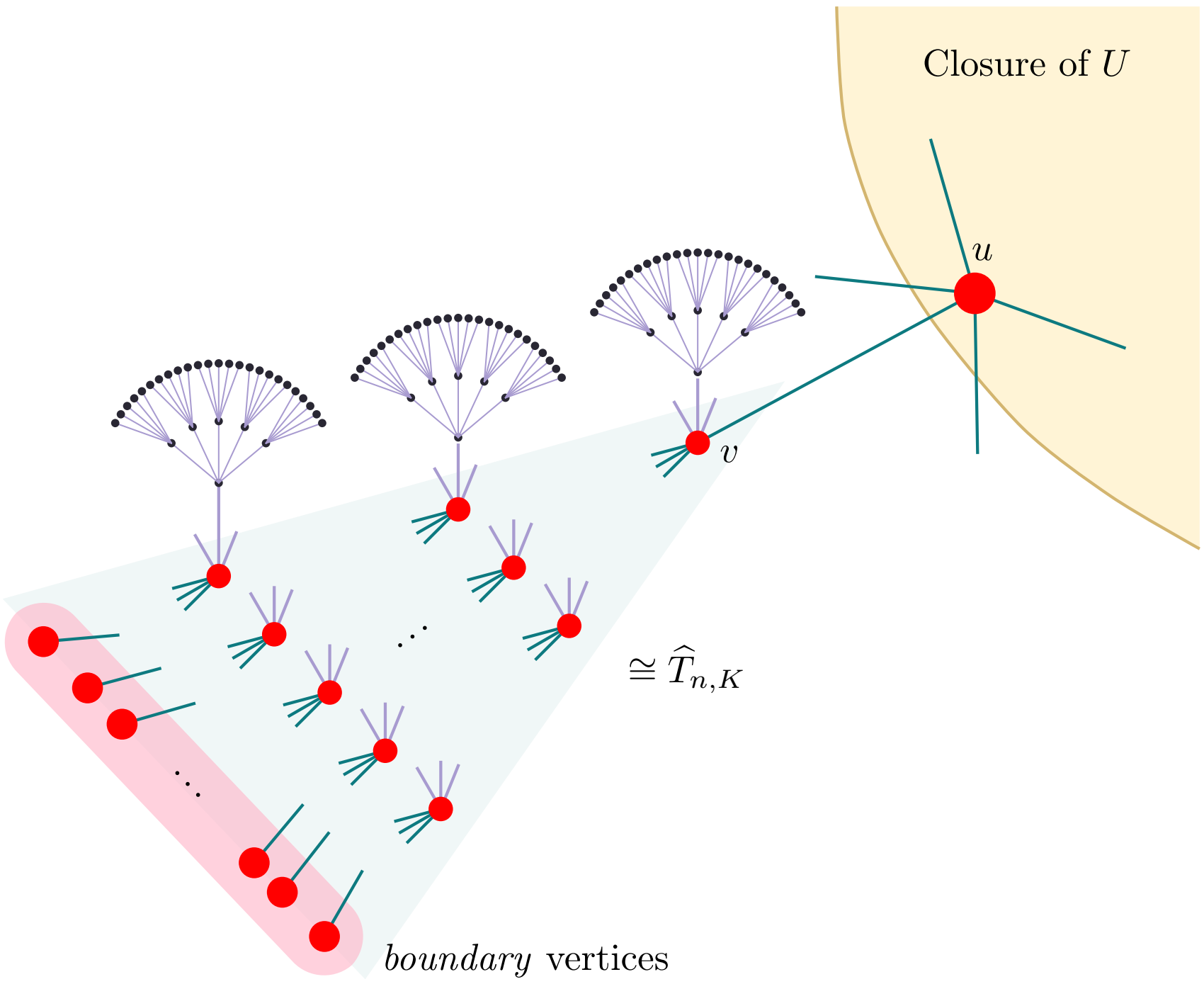}
    \caption{Pictorial representation of the subgraph reachable from a vertex $u$ in the closure of $U$ via an edge $uv$ with $v$ outside the closure. The green triangle depicts the base vertices, with their green branches indicating the $n$ base-graph neighbors, and the purple branches attached to them depict copies of self-similar trees. No such trees are attached to the farthest base vertices, which form the boundary. Together, this subgraph is isomorphic to $\treh_{n,K}$, as proved in \Cref{prop:iso}.}
  \label{Fig:baselocal}
  \end{center}
\end{figure}

\begin{proof}
  The neighbors of vertices in $W\setminus\partial U$ are contained in $W$ by step~2 of \Cref{Def:closuregraph}. The claim about the same neighborhood and~(a) follow from step~2 and~3 of \Cref{Def:closuregraph}, and so does~(c). For~(b), step~2 does not put the tree vertices attached to a boundary vertex into~$W$, and its base neighbors at distance $\ell_{n,K}+2$ from $\br{U}$ are not in~$W$ either, so $N_{\envl(U)}(v)=N_{\main}(v)\cap W\subsetneq N_{\main}(v)$ for every $v\in\partial U$.

  We now prove the second part of the claim. Consider any vertex $u \in \br{U}$. Let $v \in V_{\base} \setminus \br{U}$ be a base graph neighbor of~$u$, and let $T(u,v)$ denote the induced subgraph of $\main[W]$ consisting of the vertices reachable from $v$ \emph{after removing} the edge~$(u,v)$. It suffices to show that the subgraphs $\set{T(u,v)}_{u,v}$ are disjoint and that $T(u,v)$ is isomorphic to $\treh_{n,K}$ for all~$u,v$.

  The breadth-first forest argument in the proof of \Cref{lem:2-degenerate} shows that every base edge of~$\main[W]$ outside~$\main[\br{U}]$ belongs to a forest rooted at $\br{U}$. Since the decorations are themselves trees attached at single vertices, each edge $(u,v)$ considered above separates $T(u,v)$ from the rest of $\main[W]$. In particular, the subgraphs $T(u,v)$ are pairwise disjoint.

  Finally, since the base graph is $n$-regular, the base vertices of $T(u,v)$ form a perfect tree of depth $\ell_{n,K}$ with exactly $n-1 = d_{n,K}-1$ children at each internal vertex. All self-similar trees attached to base vertices at distance less than $\ell_{n,K}$ from $v$ are retained in $W$, whereas the base vertices at distance $\ell_{n,K}$ are exactly the boundary vertices and retain no decorations. This matches the definition of $\treh_{n,K}$ in \Cref{Def:selfsim}, with the boundary vertices corresponding to its $0$-level leaves. See \Cref{Fig:baselocal} for an illustration.
\end{proof}

We can restrict attention to algorithms that explore the graph locally, meaning that the queried vertices form connected components rooted at the vertices of $U$, together with some isolated vertices. Indeed, the proportion of non-isolated vertices in $\main$ is exponentially small, so if a classical algorithm that starts from $U$ queries a vertex that is not connected to the vertices it has already explored, this vertex is isolated in $\main$ with overwhelming probability. In particular, the algorithm queries a vertex that is isolated only in $\envl(U)$ with negligible probability, unless it reaches this vertex by exploring from $U$.

\begin{lemma}[\sc Local exploration, adapted from~\cite{HLBS26p}]\label{lem:conncomp}
  Let $U \subseteq V$ be a subset of vertices of $\main$ with $\abs{U} \leq \abs{V}/2$ and a fixed labeling, and let $G \in \set{\main,\envl(U)}$. Let $\mcl{A}$ be a classical algorithm that is given $\pi(U)$ and makes $q$ queries to the adjacency-list oracle $\ora_{G,\pi}$, for a uniformly random $\pi$ subject to the fixed labels of $U$. Then, except with probability at most $q \cdot 2^{-2n^{3}\log^2{n}}$, the set of all vertices queried by $\mcl{A}$ at any point during its execution induces a collection of connected components rooted at the vertices in~$U$, together with some vertices that are isolated in $\main$ (and hence also in $\envl(U)$).
\end{lemma}

The proof follows the same argument as in~\cite{CCD03c} or~\cite{HLBS26p}, so we omit it. Thus, we can now focus on the hardness of reaching a boundary vertex starting from $U$. By \Cref{prop:iso}, any path from a vertex in $U$ to a boundary vertex must use an edge joining a vertex in the closure to a vertex outside it, and hence contains a root-to-leaf path in a copy of the self-similar tree $\treh_{n,K}$. We use the following result on the hardness of exploring self-similar trees, first shown in~\cite{GHV21c}, in the variant suited to this construction from~\cite{HLBS26p}.

\begin{lemma}[\sc Hardness of exploring self-similar trees,~\cite{HLBS26p}]
  \label{Lem:hardss}
  Let $K = \sqrt{n}$, as in \Cref{Def:maingraph}. Suppose a classical algorithm explores the tree $\treh_{n,K}$ under a uniformly random labeling, starting at its root and querying only vertices adjacent to those it has already visited. If the algorithm makes fewer than~$2^{\sqrt{n}}$ queries, the probability that it ever reaches a $0$-level leaf of~$\treh_{n,K}$ is at most $2^{-2n\log{n}}$.
\end{lemma}

We are now ready to prove indistinguishability against any classical randomised algorithm making few queries.

\begin{theorem}[\sc Classical hardness of distinguishing]
    \label{thm:class-hard}
    Let $U$ be a set of $|U| \le 2^{\sqrt{n}}$ vertices of $\main$, and fix any partial labeling $\sigma:U \to [2^m]$. Let $\pi$ be a uniformly random labeling subject to $\pi|_U=\sigma$. Let $\mcl{A}$ be any classical randomised algorithm that is given $\pi(U)$ and makes at most $q < 2^{\sqrt{n}}$ queries to the adjacency-list oracle $\ora_{G,\pi}$, where $G \in \set{\main,\envl(U)}$. Then
    \[
        \big|\pr{\mcl{A}^{\ora_{\main,\pi}}(\pi(U)) = 1} - \pr{\mcl{A}^{\ora_{\envl(U),\pi}}(\pi(U)) = 1}\big| \leq 2^{-n\log{n}}.
    \]
\end{theorem}

\begin{proof}
  By \Cref{prop:iso}, $\main$ and $\envl(U)$ differ only on the boundary $\partial U$ and on the set $Y$. Vertices of $Y$ are not isolated in $\main$, so by \Cref{lem:conncomp}, except with probability at most $q\cdot 2^{-2n^{3}\log^2 n}$, $\mcl{A}$ queries a vertex of $Y$ only after exploring to it from $U$.
  By \Cref{prop:iso}(a), this requires first exploring a vertex of $\partial U$. Hence the advantage of $\mcl{A}$ is at most
  \[
    q\cdot 2^{-2n^{3}\log^2 n}+\pr{\mcl{A}\text{ explores a vertex of }\partial U}.
  \]
  By \Cref{prop:iso}, the vertices of $\partial U$ correspond exactly to the $0$-level leaves of the at most $n\abs{\br{U}}$ copies of $\treh_{n,K}$ attached to $\br{U}$, one for each base-graph edge leaving the closure; these copies are pairwise disjoint and independently labelled. Fix one such copy $T$. Although $\mcl{A}$'s queries to $T$ may be interleaved adaptively with its queries elsewhere, we may simulate every query of $\mcl{A}$ outside $T$ using the rest of the graph together with the given labels $\pi(U)$, which expose no internal vertex of $T$, and only charge to $T$ the at most $q$ queries that $\mcl{A}$ makes inside it. Since $q < 2^{\sqrt n}$, \Cref{Lem:hardss} applied to this restricted exploration bounds the probability that $\mcl{A}$ reaches a $0$-level leaf of $T$ by $2^{-2n\log n}$, independently of what $\mcl{A}$ learned from the other copies. A union bound over the at most $n\abs{\br{U}}$ copies, together with $\abs{\br{U}}\le 2(2\ell_{n,K}+3)\abs{U}$ (\Cref{lem:closure-size}), then gives
  \[
    \pr{\mcl{A}\text{ explores a vertex of }\partial U}\;\le\; n\abs{\br{U}}\,2^{-2n\log n}\;\le\; 2(2\ell_{n,K}+3)\,n\,2^{\sqrt n-2n\log n}.
  \]
  Using the definition of $\ell_{n,K}$, the two error terms together are at most $2^{-n\log n}$.
\end{proof}

\section{Lower bound when advice is arbitrary}
The previous section shows that polynomially many classical queries cannot distinguish $\main$ from~$\envl(U)$, even when the labels of the seed set $U$ are given as input. An $\MA$ witness, however, may be any polynomial-length bit string depending on the entire labelled graph. Since completeness only requires the existence of \emph{some} accepted witness for each labeling, we may fix, without loss of generality, one such witness $w_\pi\in\{0,1\}^{p(m)}$ for every labeling $\pi$, for some polynomial $p$. To handle such witnesses, we use a presampling lemma for random permutations from the cryptographic literature~\cite{CDG18c,CDGS18c}.

Informally, the following lemma says that an algorithm that receives a $p(m)$-bit string $w_\pi$ depending on the whole labeling $\pi$, and then queries $\pi$, can be simulated, up to a small error, by an algorithm that receives the same string but queries a labeling that is \emph{uniformly random} subject only to the labels of some $t$ vertices being fixed. In other words, the $p(m)$-bit string reveals no more about $\pi$ than the labels of $t$ vertices, chosen depending on the string. More precisely, consider the following experiment, which draws $\pi,\pi'$ jointly:
\begin{enumerate}
    \item draw $\pi$ uniformly at random among all bijections $\pi:V\to[2^m]$ (labelings of $V$), and let~$w_\pi$ be its witness;
    \item draw a partial labeling $\sigma:U\to[2^m]$, with $\abs U\le t$, from a distribution $\mcl D_{w_\pi}$ depending on~$w_\pi$;
    \item draw $\pi'$ uniformly at random among all bijections $\pi':V\to[2^m]$ subject to $\pi'|_{U}=\sigma$.
\end{enumerate}
The lemma states that querying $\pi$ on input $w_\pi$, and querying $\pi'$ on the same input $w_\pi$, give nearly the same output distribution.

\begin{lemma}[\sc Arbitrary witness to label-fixing, adapted from~\cite{CDG18c}]\label{lem:lab-fixing}
    Consider $\pi\mapsto w_\pi\in\{0,1\}^{p(m)}$, an arbitrary map, defined on labelings $\pi$ of a vertex set $V$ with $\abs{V}=2^m\ge16$. For any $1\ge\gamma>0$ and $t\in\mathbb N$, there is a family $\{\mcl D_w\}_{w\in\{0,1\}^{p(m)}}$ of distributions over partial labelings $\sigma:U\to[2^m]$ with $\abs U\le t$, such that for $\pi,\pi'$ drawn as in the experiment above, and for any classical algorithm $\mcl{B}$ that takes a $p(m)$-bit input, outputs a bit, and makes at most $q$ queries to the labeling, each either a forward query $v\mapsto\pi(v)$ or a backward query $x\mapsto\pi^{-1}(x)$,
    \[
        \abs[\Big]{\pr_{\pi}[\big]{\mcl{B}^{\pi}(w_\pi)=1}
        -\pr_{\pi,\pi'}[\big]{\mcl{B}^{\pi'}(w_\pi)=1}}
        \le \frac{2\,(p(m)+\log 1/\gamma)\cdot q}{t}+\gamma.
    \]
\end{lemma}

In our case, the algorithm queries the relabelled adjacency-list oracle $\ora_{G,\pi}$ rather than the labeling $\pi$ directly. The corollary below shows that an algorithm querying $\pi$ can simulate one querying $\ora_{G,\pi}$, at a cost of a factor $2n$ in the number of queries. Hence \Cref{lem:lab-fixing} applies, and the witness can again be treated as essentially only a set of vertices with fixed labels.

\begin{corollary}[\sc Simulating adjacency-list queries by labeling queries]
    \label{cor:adj-to-perm}
    Let $G\in\{\main,\envl(U)\}$ for some set $U$.
    For $p,\gamma,t$ and $\pi,\pi'$ as in \Cref{lem:lab-fixing}, and every classical algorithm~$\mcl{A}$ that takes a $p(m)$-bit input and makes at most $q$ queries to the adjacency-list oracle $\ora_{G,\pi}$ (\Cref{Def:ora-graph}),
    \[
        \abs[\Big]{\pr_{\pi}[\big]{\mcl{A}^{\ora_{G,\pi}}(w_\pi)=1}
        -\pr_{\pi,\pi'}[\big]{\mcl{A}^{\ora_{G,\pi'}}(w_\pi)=1}}
        \le \frac{2\,(p(m)+\log 1/\gamma)\cdot 2nq}{t}+\gamma.
    \]
\end{corollary}

\begin{proof}
    There is an algorithm $\mcl{B}$ that simulates $\mcl{A}$ using only labeling queries. $\mcl{B}$ has $G$ hard-wired, so it can look up neighborhoods in $G$ for free. It runs $\mcl{A}$ and answers each query $(x,i)$ as follows: it computes $u=\pi^{-1}(x)$ with one backward query, then queries $\pi(v)$ for every neighbor $v$ of $u$ in $G$, sorts these labels, and returns the $i$th one (or $x$ if $i>\deg_G(u)$). This gives exactly the answer of $\ora_{G,\pi}$, so $\mcl{B}^{\pi}(w)$ and $\mcl{A}^{\ora_{G,\pi}}(w)$ have the same output distribution for every labeling $\pi$ and every input $w$. In particular,
    \[
        \pr{\mcl{B}^{\pi}(w_\pi)=1}=\pr{\mcl{A}^{\ora_{G,\pi}}(w_\pi)=1}
        \quad\text{and}\quad
        \pr{\mcl{B}^{\pi'}(w_\pi)=1}=\pr{\mcl{A}^{\ora_{G,\pi'}}(w_\pi)=1}.
    \]
    Since $\main$, hence also $\envl(U)$, has maximum degree less than $2n$, each query of $\mcl{A}$ costs $\mcl{B}$ at most $1+\deg_G(u)\le 2n$ queries, so $\mcl{B}$ makes at most $2nq$ queries.
    The family $\{\mcl D_w\}$ from \Cref{lem:lab-fixing} does not depend on the algorithm, so applying that lemma to $\mcl{B}$, with $2nq$ in place of $q$, and substituting the two equalities above gives the claimed bound.
\end{proof}

We now combine this corollary with \Cref{thm:class-hard} to show that arbitrary polynomial-length witnesses cannot help distinguish the two graph families.

\begin{proposition}
    \label{lem:arb-witness}
    There is no $\MA$ protocol that solves the GPE problem using polynomial witness length and polynomially many queries to $\ora_{G,\pi}$, where $G\in\{\main,\envl(U)\}$ and $U$ ranges over all sets of at most $2^{\sqrt{n}}$ vertices. In fact, any such protocol errs on some instance with parameter $n$ for every sufficiently large $n$.
\end{proposition}

\begin{proof}
    Fix a protocol with a verifier $\mcl{V}$ that makes at most $q\le\poly(n)$ queries and uses witnesses of length at most $p(m)=\poly(n)$, and suppose that it solves the problem on all instances with parameter $n$, for a sufficiently large $n$. For every $\pi$ it satisfies:
    \begin{itemize}
        \item \emph{Completeness.} There is a witness $w\in\rn^{p(m)}$ with $\pr{\mcl{V}^{\ora_{\main,\pi}}(w)=1}\ge 2/3$. Fix one such witness $w_\pi$ for every $\pi$.
        \item \emph{Soundness.} For every $U$ with $\abs{U}\le 2^{\sqrt{n}}$ and every witness $w$, $\pr{\mcl{V}^{\ora_{\envl(U),\pi}}(w)=1}\le 1/3$.
    \end{itemize}
    Let $t=2^{\sqrt{n}}$ and $\gamma=2^{-n}$. Applying \Cref{cor:adj-to-perm,lem:lab-fixing} to $\mcl V^{\ora_{\main,\pi}}$, for large enough $n$, the experiment of \Cref{lem:lab-fixing} gives
    \[
        \left|\pr_{\pi}*{\mcl{V}^{\ora_{\main,\pi}}(w_\pi)=1}
        -\pr_{\pi,\pi'}*{\mcl{V}^{\ora_{\main,\pi'}}(w_\pi)=1}\right|
        \le \frac{2(p(m)+n)\cdot 2nq}{2^{\sqrt{n}}}+2^{-n}\le 2^{-\sqrt{n}/2}.
    \]
    By completeness the first probability is at least $2/3$, so the second is at least $2/3-2^{-\sqrt{n}/2}$. The second probability is an average over $(\pi,\sigma)$, so there are a $\pi$ and a $\sigma$ in the support of $\mcl D_{w_\pi}$ such that, writing $w=w_\pi$ and $U=\operatorname{dom}\sigma$ (so $\abs{U}\le t$), and with $\pi'$ uniform among labelings with $\pi'|_U=\sigma$ (as in the rest of the proof),
    \begin{equation}\label{eq:yes-fixed}
        \pr_{\pi'}*{\mcl{V}^{\ora_{\main,\pi'}}(w)=1}\ge \frac23-2^{-\sqrt{n}/2}.
    \end{equation}

    The witness $w$ is fixed before $\pi'$ is drawn, so it is a constant. Let $\mcl{A}_w$ be the algorithm that has $w$ hard-wired, takes $\sigma$ as input, and runs $\mcl{V}(w)$. Then $\mcl{A}_w^{\ora_{G,\pi'}}(\sigma)$ and $\mcl{V}^{\ora_{G,\pi'}}(w)$ have the same output distribution for both $G\in\{\main,\envl(U)\}$. Since $\sigma$ is exactly the fixed assignment in the hypothesis of \Cref{thm:class-hard}, that theorem applies to $\mcl{A}_w$:
    \[
        \begin{array}{l}
            \left|\pr_{\pi'}*{\mcl{V}^{\ora_{\main,\pi'}}(w)=1}
            -\pr_{\pi'}*{\mcl{V}^{\ora_{\envl(U),\pi'}}(w)=1}\right|
            \\
            =\left|\pr_{\pi'}*{\mcl{A}_w^{\ora_{\main,\pi'}}(\sigma)=1}
            -\pr_{\pi'}*{\mcl{A}_w^{\ora_{\envl(U),\pi'}}(\sigma)=1}\right|
            \le 2^{-n\log n}.
        \end{array}
    \]
    Together with \eqref{eq:yes-fixed} this gives
    \[
        \pr_{\pi'}*{\mcl{V}^{\ora_{\envl(U),\pi'}}(w)=1}\ge \tfrac23-2^{-\sqrt{n}/2}-2^{-n\log n}>\tfrac13 .
    \]
    So some $\pi'$ with $\pi'|_U=\sigma$ makes $\mcl{V}$ accept the $\no$ instance $\envl(U)$ with witness $w$ with probability more than $1/3$. This violates soundness, a contradiction.
\end{proof}

We have shown an \emph{oracle-input} separation for GPE: the graph oracle is itself the instance, and a protocol given a witness and access to it, with no separate input string, decides whether it is a $\yes$ or $\no$ instance. Thus GPE belongs to $\StoqMA$ but not to $\MA$ in this model. The remaining task is to turn this into a genuine separation of the relativized classes $\StoqMA^{\mcl{O}}$ and $\MA^{\mcl{O}}$ for an explicit oracle~$\mcl{O}$. This is a standard diagonalization argument, proved formally in~\cite{Zha25c} and used in several earlier separations; we include a short proof adapted to our setting for completeness.

\begin{rtheorem}[\Cref{thm:main}]
    There exists a classical oracle $\mcl{O}$  such that $\MA^{\mcl{O}} \subsetneq \StoqMA^{\mcl{O}}$.
\end{rtheorem}

\begin{proof}
    We will design an oracle $\mcl{O}$ and a language $\mcl{L}^{\mcl{O}}$ such that $\mcl{L}^{\mcl{O}} \in \StoqMA^{\mcl{O}}$ but $\mcl{L}^{\mcl{O}} \notin \MA^{\mcl{O}}$. The oracle will be a sequence of graph oracles from the universe defined below.

    For each $n$, let $m = m(n) = 15n^3\log^2 n$ and $d = 2n-\sqrt{n}$. The universe is $\mcl{U}=\{\mcl{U}_n\}_{n\in\mathbb{Z}_+}$, where $\mcl{U}_n$ is the set of adjacency-list oracles $\ora_{G,\pi}\colon[2^m]\times[d]\to[2^m]$ (\Cref{Def:ora-graph}), one for each graph $G\in\{G_n\}\cup\{F_n(U): U\subseteq V(G_n),\ \abs{U}\le 2^{\sqrt n}\}$ and each bijective labeling $\pi\colon V(G)\to[2^m]$. As noted after \Cref{def:stoqma}, these oracles fit $\MA^{\mcl{O}}$ and $\StoqMA^{\mcl{O}}$ via the one-bit-at-a-time simulation, at only polynomial overhead.

    We build $\mcl{O}$ in stages, diagonalising against all candidate $\MA$ verifiers. The oracle is a family of blocks $\{\ora^{n}\}_{n\in\mathbb{Z}_+}$ with $\ora^{n}\in\mcl{U}_n$. Initially every block is $\ora_{F_{n}(\emptyset),\mathrm{id}}$, where $\mathrm{id}$ is the identity permutation. At stage $i$ we choose an integer $n_i>n_{i-1}$ and replace block $n_i$ by an instance $\ora^{n_i}\in\mcl{U}_{n_i}$, chosen below, while every block $n\notin\{n_i\}$ stays $\ora_{F_{n}(\emptyset),\mathrm{id}}$. Let $\mcl{L}^{\mcl{O}}_{\rm yes}=\{0^n:\ora^{n}=\ora_{G_n,\pi}\text{ for some }\pi\}$ and $\mcl{L}^{\mcl{O}}_{\rm no}=\{0,1\}^{\star}\setminus\mcl{L}^{\mcl{O}}_{\rm yes}$.

    As in~\cite{Zha25c}, we pad all witnesses with $0$'s so that any potential $\MA$ verifier runs in quadratic time. Let $\mcl{V}_1,\mcl{V}_2,\dots$ enumerate the quadratic-time randomised verifiers $\mcl{V}_i^{\mcl{O}}(x,w)$ with oracle access as in~\Cref{def:ma}. At stage $i$ we want an input $0^{n_i}$ on which $\mcl{V}_i$ errs. Our lower bound, however, is for protocols whose only input is the witness and which query a single graph oracle. We therefore derive such a protocol from $\mcl{V}_i$. Let $\mcl{O}_{i-1}$ be the oracle with blocks $n_1,\dots,n_{i-1}$ chosen and all other blocks $\ora_{F_{n}(\emptyset),\mathrm{id}}$. For an instance $\ora_{G,\pi}\in\mcl{U}_n$, let $\widetilde{\mcl{V}}_i^{\ora_{G,\pi}}(w)$ simulate $\mcl{V}_i(0^n,w)$ with the oracle $\mcl{O}_{i-1}[n\leftarrow\ora_{G,\pi}]$, in which block $n$ is replaced by $\ora_{G,\pi}$. It hard-codes the blocks $n_1,\dots,n_{i-1}$, whose size is independent of $n$ and $w$, so $\widetilde{\mcl{V}}_i$ runs in polynomial time, up to a constant depending on $i$. It is an $\MA$ protocol of the kind covered by~\Cref{lem:arb-witness}.

    We choose $n_i$ and $\ora^{n_i}\in\mcl{U}_{n_i}$ such that both of the following hold:
    \begin{itemize}
        \item $\widetilde{\mcl{V}}_i^{\ora^{n_i}}(w)$ decides $\ora^{n_i}$ incorrectly: either $\ora^{n_i}=\ora_{G_{n_i},\pi}$ for some $\pi$ and every witness $w$ is accepted with probability less than $2/3$, or $\ora^{n_i}=\ora_{F_{n_i}(U),\pi}$ and some witness is accepted with probability more than $1/3$.
        \item $n_i>n_{i-1}$ is large enough that the queries to block $n_i$ are longer than the running time of $\mcl{V}_j$ on inputs $(0^{n_j},w)$, over all witnesses $w$ of the padded length, for all $j<i$.
    \end{itemize}
    Such $n_i$ and $\ora^{n_i}$ exist because the lower bound of~\Cref{lem:arb-witness} holds for every sufficiently large $n$ and applies to $\widetilde{\mcl{V}}_i$.

    We claim that $\mcl{L}^{\mcl{O}}=\mcl{L}^{\mcl{O}}_{\rm yes}$ is in $\StoqMA^{\mcl{O}}$. For strings other than $0^n$ we reject by a simple check. For $0^n$ we run the $\StoqMA$ protocol of~\Cref{prop:gpe-stoqma} on block $\ora^{n}$, which decides whether $0^n$ is a $\yes$ instance.

    We now show that $\mcl{L}^{\mcl{O}}\notin\MA^{\mcl{O}}$. Suppose not, and let $\mcl{V}$ be an $\MA$ verifier for it. After padding it runs in quadratic time, so $\mcl{V}=\mcl{V}_i$ for some $i$ in our enumeration. On input $(0^{n_i},w)$, $\mcl{V}_i$ cannot query the blocks $n_j$ for $j>i$, by the choice of $n_j$, and the blocks that are not chosen are the fixed empty-graph blocks. So it behaves exactly as $\mcl{V}_i(0^{n_i},w)$ with oracle $\mcl{O}_{i-1}[n_i\leftarrow\ora^{n_i}]$, which is $\widetilde{\mcl{V}}_i^{\ora^{n_i}}(w)$ by definition. But $\widetilde{\mcl{V}}_i^{\ora^{n_i}}$ decides $\ora^{n_i}$ incorrectly, so $\mcl{V}_i$ decides $0^{n_i}$ incorrectly. This contradicts $\mcl{V}$ being a valid verifier for $\mcl{L}^{\mcl{O}}$.
\end{proof}

\section*{Acknowledgements}
We used AI (ChatGPT 5.6) during our search for a suitable NO-instance construction. It helped us identify an $\NP$ protocol for a previous construction studied in~\cite{HLBS26p} (see \Cref{App:earlier-problem-np}) and pointed out that another of our earlier candidate graphs was 2-degenerate, suggesting a structural property that could be exploited to construct an alternative NO instance. This observation motivated our subsequent use of the closure and 2-degeneracy, which we developed ourselves for the final family of NO instances. We also used AI-based tools to assist with writing the TikZ code for the figures, as well as for language editing and proofreading.

This work was supported by the Maison du Quantique de Nouvelle-Aquitaine ``HybQuant'', as part of the HQI initiative and France 2030, under the French National Research Agency (ANR) grant ANR-22-PNCQ-0002. It was also supported by the PEPR integrated project EPiQ under ANR grant ANR-22-PETQ-0007.

\printbibliography[heading=bibintoc]

\appendix
\section{An in-place oracle separation for \texorpdfstring{$\StoqMA$}{StoqMA}}
\label[appendix]{App:preimage-stoqma}

In the Preimage Checking problem~\cite{FK18c}, an unknown set $S \subseteq [N^2]$ has size $\abs{S} = N$. Writing $S_{\mathrm{even}}$ for its even elements, the promise is $\yes: \abs{S_{\mathrm{even}}} = 2N/3$, $\no: \abs{S_{\mathrm{even}}} = N/3$. The in-place oracle applies $\ora_\pi\ket{x} = \ket{\pi(x)}$ for a permutation $\pi$ satisfying $\pi(S) = [N]$.

\begin{proposition}
  \label{Prop:preimage-stoqma}
  The Preimage Checking problem has a one-query stoquastic verifier with completeness at least $5/6$ and soundness at most $2/3$.
\end{proposition}

\begin{proof}
  The witness is assumed to be a superposition over $[N^2]$, written $\ket{\psi} = \sum_{x\in[N^2]} \beta_x\ket{x}$. The verifier computes the parity into a fresh qubit and applies the oracle to the original register, obtaining $\sum_x \beta_x\ket{\pi(x)}\ket{x\bmod 2}$. It then performs a SWAP test with the state $(\frac{1}{\sqrt N}\sum_{y\in[N]}\ket{y})\ket{0}$.

  The acceptance probability is $p_{\mathrm{acc}} = \frac{1}{2} + \frac{1}{2N}\abs{\sum_{x\in S_{\mathrm{even}}}\beta_x}^2 \leq \frac{1}{2} + \frac{\abs{S_{\mathrm{even}}}}{2N}$, by the Cauchy--Schwarz inequality. The uniform witness $\ket{S_{\mathrm{even}}}$ attains equality in the YES case. The promised cardinalities give completeness at least $5/6$ and soundness at most $2/3$.
\end{proof}

Fefferman and Kimmel~\cite{FK18c} proved a $\QCMA$ lower bound for Preimage Checking in a randomized oracle model in which each query uses a fresh uniformly random permutation $\pi$ satisfying $\pi(S) = [N]$. This per-query randomness was recently removed for a closely related preimage-parity problem~\cite{MPR26c}, yielding $\QMA_1^{\ora} \not\subseteq \QCMA^{\ora}$ for a fixed in-place permutation oracle $\ora$. The verifier above also applies to the promise in~\cite{MPR26c}, with completeness $1$ and soundness at most $3/4$. Combined with the $\QCMA$ lower bound established in that work, this yields $\StoqMA_1^{\ora} \not\subseteq \QCMA^{\ora}$.

This result does not contradict the (unrelativized) containment $\StoqMA_1 \subseteq \MA$ shown in~\cite{AGL25j}. The proof of that containment requires inverting the verifier's circuit, which is generally not possible with an in-place oracle. This separation also says nothing about $\StoqMA$ vs. $\MA$ relative to classical (XOR) oracles.


\section{An \texorpdfstring{$\NP$}{NP} protocol for the earlier graph problem}
\label[appendix]{App:earlier-problem-np}

The graph-testing problem underlying the lower bound in our previous work~\cite{HLBS26p} is to distinguish two graphs $\mathcal{G}_n$ and $\mathcal{F}_n$. The graph $\mathcal{G}_n$ is constructed as $\main$ in the present paper. The graph $\mathcal{F}_n$ uses a base graph of degree $n-\sqrt n$ instead of $n$, with additional tree decorations ensuring that both graphs have the same degree at every non-isolated, non-leaf vertex. We show that this problem lies in $\NP$, although~\cite{HLBS26p} proves it hard for classical algorithms given samples from a guiding state.

\begin{proposition}
  \label{Prop:np-protocol}
  The $\mathcal{G}_n$ (YES) versus $\mathcal{F}_n$ (NO) problem defined in~\cite{HLBS26p} is in $\NP$.
\end{proposition}

We use the following general fact: in any graph of minimum degree $n$, some vertex is incident to at least $n-2$ distinct edges lying on short cycles.

\begin{lemma}
  \label{Lem:spanner}
  Let $E$ be a graph on $N$ vertices with minimum degree at least $n \geq 2$. Some vertex is incident to at least $n-2$ distinct edges, each lying on a cycle of length at most $2\ceil{\log N}+2$.
\end{lemma}

\begin{proof}
  Set $k = \ceil{\log N}+1$. Construct a greedy spanner $J$: starting with no edges, process each edge of $E$, adding it only when its endpoints currently have distance greater than $2k-1$. Then~$J$ has girth greater than $2k$. If every vertex of $J$ had degree at least three, a breadth-first tree of depth $k$ would contain at least $1+3(2^k-1) > N$ vertices, since the girth bound prevents collisions. Thus some vertex $v$ has degree at most two in $J$. At least $n-2$ edges incident to~$v$ were omitted, each with an alternative path of length at most $2k-1$, giving the claimed cycles.
\end{proof}

\begin{proof}[Proof sketch of Proposition~\ref{Prop:np-protocol}]
  Let $N$ be the number of vertices in the base graph of $\mathcal{G}_n$. By Lemma~\ref{Lem:spanner}, this base graph has a vertex incident to $n-2$ distinct edges, each lying on a cycle of length $\bo{\log N}$. The prover supplies this vertex, these edges, and their cycles; the verifier checks the certificate using adjacency queries. In $\mathcal{F}_n$, each vertex is incident to at most $n-\sqrt n < n-2$ edges lying on cycles, proving soundness.
\end{proof}

\end{document}